\documentclass[a4paper, leqno, 12pt]{article}
\usepackage{amsmath,amsthm}
\usepackage{amssymb,latexsym}
\usepackage{enumerate}
\usepackage[T1]{fontenc}
\usepackage{array}
\usepackage[all]{xypic}

\usepackage{tikz}
\usetikzlibrary{decorations.pathmorphing}
\usetikzlibrary{shapes,decorations.markings,decorations.pathreplacing,backgrounds,positioning,calc}

\usepackage[left=3cm,right=3cm,top=3.0cm,bottom=3.0cm]{geometry}

\def\CCC{{\mathbb{C}}}

\def\NN{{\mathbb{N}}}
\def\MM{{\mathbb{M}}}

\def\RR{{\mathbb{R}}}
\def\TT{{\mathbb{T}}}

\def\CB{{\cal B}}

\def\CF{{\cal F}}

\def\CH{{\cal H}}

\def\CL{{\cal L}}

\def\CR{{\cal R}}
\def\CS{{\cal S}}
\def\CZ{{\cal Z}}

\def\CZ{{\cal Z}}
\def\CV{{\cal V}}

\def\rem{\textnormal{rem}} 
\def\sgn{\textnormal{sgn}} 
\def\snull{\textnormal{ }}

\def\epv {{$\mbox{}$\hfill ${\Box}$\vspace*{1.5ex} }}

\newcommand{\ra}{\rightarrow}

\newcommand{\Ra}{\Rightarrow}
\newcommand{\Lra}{\Leftrightarrow}

\def\ov{\overline} 
\def\wt{\widetilde}
\def\wh{\widehat}
\def\epv {{$\mbox{}$\hfill ${\Box}$\vspace*{1.5ex} }}

\newtheorem{thm}{Theorem}[section]
\newtheorem{cor}[thm]{Corollary}

\newtheorem{prop}[thm]{Proposition}

\theoremstyle{definition}
\newtheorem{df}[thm]{Definition}
\newtheorem{alg}[thm]{Algorithm}
\newtheorem{ex}[thm]{Example}
\newtheorem{remark}[thm]{Remark}
\newtheorem{cons}[thm]{Construction}

\title{\textmd{{\Large On maps which preserve semipositivity and quantifier elimination theory for real numbers}}}
\date{}
\begin{document}
\maketitle
\begin{center}
Grzegorz Pastuszak$^{a}$\footnote{Corresponding author}, Adam Skowyrski$^{a}$ and Andrzej Jamio{\l}kowski$^{b}$\\\textnormal{ }\\
\footnotesize{$^{a}$ Faculty of Mathematics and Computer Science, Nicolaus Copernicus University, Toru\'n, Poland, past@mat.umk.pl (Grzegorz Pastuszak), skowyr@mat.umk.pl (Adam Skowyrski); $^{b}$ Faculty of Physics, Astronomy and Informatics, Nicolaus Copernicus University, Toru\'n, Poland, jam@fizyka.umk.pl}
\end{center}

\begin{abstract}
Assume that $\Phi:\MM_{n}(\CCC)\ra\MM_{n}(\CCC)$ is a superoperator which preserves hermiticity. We give an algorithm determining whether $\Phi$ preserves semipositivity (we call $\Phi$ \textit{positive} in this case). Our approach to the problem has a model-theoretic nature, namely, we apply techniques of quantifier elimination theory for real numbers. An approach based on these techniques seems to be the only one that allows to decide whether an arbitrary hermiticity-preserving $\Phi$ is positive. Before we go to detailed analysis of the problem, we argue that quantifier elimination for real numbers (and also for complex numbers) can play a significant role in quantum information theory and other areas as well.  
\end{abstract}

{\footnotesize}

\section{Introduction and the main results} 

The dynamics of a finite isolated quantum system is usually described by a one-parameter group of unitary transformations in a complex Hilbert space (cf. e.g. \cite{BZ}). However, in many physical problems it is necessary to consider a given quantum system as an open one which interacts with its surroundings. 

In modelling of open systems for which time behaviour can be represented by stochastic processes, one assumes that a system in question is described by certain mathematical model, for example by random variables in the classical case or by sets of non-commuting observables in the quantum case, acting on an abstract probability space. In algebraic formulation of quantum mechanics, a fixed quantum mechanical system is represented by an algebra $A$ of operators acting on some Hilbert space $\CH$. In this approach, the observables (i.e. measured quantities) of the system are identified with hermitian (i.e. selfadjoint) elements in $A$ and physical states are given by the set $S(\CH)$ of density operators, that is, semipositive elements in $A$ with unital trace. Evolutions of the system are described by maps on the set $S(\CH)$. This means that we are interested in \textit{positive} maps, that is, maps sending semipositive operators to semipositive operators. Such maps are superoperators of some particular form. 

The general form of superoperators which preserve hermiticity of operators is well  known, see \cite{HZ} or Section 3. An important problem of finding among such superoperators those which preserve semipositivity is still an open one. In this paper we address the problem by applying techniques of quantifier elimination theory for real numbers.

Quantifier elimination is a concept that appears in a field of mathematical logic called \textit{model theory}. It is especially important in the \textit{first-order logic} which is roughly the same as the \textit{predicate calculus}. Informally, quantifier elimination, if possible, allows to associate with a first-order formula $\varphi$ a quantifier-free formula $\varphi'$ in such a way that these two formulas are equivalent. Recall that a formula $\psi$ is \textit{quantifier-free} if and only if quantifiers $\exists$ and $\forall$ do not occur in $\psi$. Therefore quantifier-free formulas are straightforward to verify, unlike in the case of general formulas. In this sense, quantifier elimination can be viewed as a method for verifying validity of complicated formulas. We refer to Section 2 for more information on first-order formulas and quantifier-free formulas. A good introduction to first-order logic and model theory is given in \cite{M} or \cite{Rot}. 

A well-known trivial example of quantifier elimination concerns the existence of a real root of a real quadratic polynomial. Indeed, consider a formula $\varphi$ of the following form $\exists_{x\snull} a\neq 0 \wedge ax^{2}+b^{2}+c=0$ where $a,b,c\in\RR$ (this formula is in fact a \textit{sentence}). Then $\varphi$ is equivalent with the quantifier-free formula $\varphi'$ of the form $a\neq 0\wedge b^{2}-4ac\geq 0$. Recall that similar conditions are known for real cubic and real quartic polynomials. More generally, consider a first-order formula $\varphi$ which states that two non-zero real polynomials $p,q\in\RR[x]$ have a common root (possibly in the field $\CCC$ of complex numbers). Then $\varphi$ is equivalent with the quantifier-free formula stating that the \textit{resultant} of $p$ and $q$ is non-zero, see \cite{Mi} or \cite{Wa} for details. As a more advanced example, consider the \textit{quartic problem} \cite{AM} which concerns finding conditions on real numbers $p,q,r$ so that $x^{4}+px^{2}+qx+r$ is a non-negative real number, for any $x\in\RR$. In other words, we consider a formula of the form $\forall_{x\snull}x^{4}+px^{2}+qx+r\geq 0$. It is proved in \cite{AM} that this formula is equivalent with the quantifier-free formula $$\delta\geq 0\wedge(p\geq 0\vee L<0\vee (L=0\wedge q=0))$$ where $L=8pr-9q^{2}-2p^{3}$ and $\delta=256r^{3}-128p^{2}r^{2}+144pq^{2}r+16p^{4}r-27q^{4}-4p^{3}q^{2}$. Note that \cite{AM} gives three other solutions to the quartic problem. We refer the reader to \cite{XY2} for similar considerations, see also \cite{Ar}, \cite{Ch} and \cite{Ya} as valuable addenda.  

In this paper we apply one of the most prominent results in model theory, known as the \textit{Tarski-Seidenberg theorem}, stating that \textit{the theory of real closed fields admits quantifier elimination}. This fundamental result is a generalization of a theorem proved by A. Tarski in 1931 on decidability of the theory of real closed fields. The Tarski-Seidenberg theorem is described in detail in Section 2. We aim to present it in accessible way which avoids abstract logical terminology. For this purpose, we concentrate on the field $\RR$ of real numbers which is the main example of a real closed field. We refer the reader to \cite{vdD} for interesting historical remarks concerning this outstanding fact, as well as its proof. 

The crucial consequence of the Tarski-Seidenberg theorem is that  we are able to eliminate quantifiers in formulas (properly) composed from equalities and inequalities of real multivariate polynomials. Importantly, this can be done in an effective way, that is, we can \textit{compute} a quantifier-free formula equivalent with the given one. This opens a possibility for applications of quantifier elimination theory in many areas of physics and applied mathematics, see Section 2 for more comprehensive discussion. 

The present paper supports the above assertion. Indeed, we study the problem of determining whether a hermiticity-preserving superoperator is a positive map. Our strategy is the following. Assume that $\Phi:\MM_{n}(\CCC)\ra\MM_{n}(\CCC)$ is a superoperator which \textit{preserves hermiticity}, that is, $\Phi(X)$ is a hermitian matrix whenever $X$ is hermitian. Recall that $\Phi$ is \textit{positive} if and only if $\Phi(X)$ is a semipositive matrix whenever $X$ is semipositive. We associate with $\Phi$ some real multivariate polynomial $p_{\Phi}$ in $4n$ variables such that $\Phi$ is positive if and only if $p_{\Phi}(a_{1},\dots,a_{4n})\geq 0$, for any $a_{1},\dots,a_{4n}\in\RR$ (this is denoted by $p_{\Phi}\geq 0$). In terms of first-order logic, the latter condition means that we consider the validity of the following first-order formula, say $\varphi$, over the field $\RR$ of real numbers: $\forall_{a_{1}}\forall_{a_{2}}\dots\forall_{a_{4n}\snull}p_{\Phi}(a_{1},\dots,a_{4n})\geq 0$. Then the Tarski-Seidenberg theorem yields the existence of a quantifier-free formula $\varphi'$ which is equivalent with $\varphi$. Since $\varphi'$ can be computed, we are able to determine the validity of $\varphi$ by determining the validity of $\varphi'$.

Note that similar approaches to the problem we consider are known, see especially \cite{Ja1,Ja2}, \cite{SkZ} and \cite{Ch}. However, this paper is the first that gives a \textit{concrete procedure}, based on techniques of quantifier elimination theory, which determines whether the formula $\varphi$ holds or not, see Algorithm 5.7. Observe that in particular we are not interested in the explicit quantifier-free form $\varphi'$ of $\varphi$. Nonetheless, our way is completely sufficient for applications (and rather close to determining $\varphi'$). 

The procedure presented in Algorithm 5.7 is based on techniques of J. Renegar given in the series of papers \cite{Re1,Re2,Re3}. To the best of our knowledge, these results provide the most straightforward approach to the generally difficult quantifier elimination for the field of real numbers. Importantly, they are also quite effective from the point of view of computational complexity. We stress that there exists a vast literature on quantifier elimination for real closed fields (and for the field of real numbers in particular). The interested reader is referred to the huge monograph \cite{BPR}, see also \cite{Mi}. Note that the formula $\varphi$ described above has a special simple form - it is a negation of an existential formula. Therefore we mainly apply the results of \cite{Re1} which deals with the \textit{existential theory} of real numbers, see the first section of \cite{Re1} for more information.

The paper is organized as follows. In Section 2 we give some basic information on first-order formulas over the field $\RR$ of real numbers. Our main purpose is to formulate the theorem of Tarski and Seidenberg in a concise manner, avoiding abstract logical notions and terminology. We also discuss its possible applications in fields of science that rely on mathematics. 

The last part of Section 2 is a brief comment on another famous result in the first-order logic, stating that \textit{the theory of algebraically closed field admits quantifier elimination}. Interestingly, this is also due to Tarski. Although we do not make any use of quantifier elimination for algebraically closed fields, we view discussing this matter as valuable and natural. Indeed, since the field $\CCC$ of complex numbers is an algebraically closed field, the potential for applications of Tarski's result is as high as in the previous case. In fact, we show a concrete example of such an application in quantum information theory. The example comes from \cite{Pa} and concerns irreducible completely positive superoperators.

We emphasize that Section 2 is intentionally designed as a bit less formal \textit{tale on quantifier elimination}. Our aim is to advertise the use of this tool rather then scare away by abstract logical formalism. The remaining Sections 3, 4 and 5, which are the core of the paper, are written with full mathematical precision. 

Section 3 is devoted to show that if $\Phi:\MM_{n}(\CCC)\ra\MM_{n}(\CCC)$ is a superoperator preserving hermiticity, then there exists some real homogeneous polynomial $p_{\Phi}$ of degree 4 in $4n$ variables, called the \textit{positivity polynomial}, such that $\Phi$ is positive if and only if $p_{\Phi}\geq 0$. This fact is already known in the literature, but in our opinion it lacks a rigorous mathematical proof. We give such a proof in Theorem 3.2. The precise form of $p_{\Phi}$ is calculated in Theorem 3.5, see in particular the assertion $(2)$. 

In Section 4 we recall the \textit{generalized Sturm's theorem}, also known as the \textit{Sturm-Tarski theorem}. Assume that $f,g\in\RR[x]$ are non-zero univariate polynomials and denote by $N(f,g)$ the following natural number: $$|\{x\in\RR\mid f(x)=0\wedge g(x)>0\}|-|\{x\in\RR\mid f(x)=0\wedge g(x)<0\}|$$ where $|X|$ is the cardinality of the set $X$. The generalized Sturm's theorem gives a method for calculating the value of $N(f,g)$. Observe that if $g\in\RR[x]$ is a polynomial such that $g(x)>0$, for any $x\in\RR$, then $N(f,g)$ is the number of distinct real roots of polynomial $f$. The Sturm-Tarski theorem has incredibly many applications, see for example \cite{BPR}. We apply it in determining the validity of the sentence $$\exists_{x\snull}(p(x)>0\wedge q(x)>0)$$ where $p,q\in\RR[x]$. This is given in Algorithm 4.4, see also Corollary 4.3. Algorithm 4.4 is directly applied in Section 5.

Section 5 is devoted to present a procedure for determining whether a homogeneous real polynomial $g\in\RR[x_{1},\dots,x_{n}]$ of an even degree satisfies the inequality $g(a_{1},\dots,a_{n})\geq 0$, for any $a_{1},\dots,a_{n}\in\RR$ (we write $g\geq 0$ for short). Recall that the positivity polynomial $p_{\Phi}$, studied in Section 3, is homogeneous of an even degree, so this setting is sufficient. The procedure is presented in Algorithm 5.6. In this algorithm we adjust the general decision method for the existential theory of real numbers given by J. Renegar in \cite{Re1}. 

Algorithm 5.6 is based on three rather technical constructions, see Constructions 5.1, 5.3 and 5.4. The aim of these constructions is to show that the formula $g\geq 0$ is equivalent with some logical condition based on sentences considered in Section 4, that is, sentences of the form $\exists_{x\snull}(p(x)>0\wedge q(x)>0)$ where $p,q\in\RR[x]$, see Theorem 5.2 for the precise statement. This allows to apply Algorithm 4.4 in Algorithm 5.6. The procedure which determines whether a hermiticity-preserving superoperator is positive is a direct consequence of Algorithm 5.6 and Theorem 3.5. We present the procedure in Algorithm 5.7.  

\section{On quantifier elimination and its applications}

This section is devoted to present the theorem of Tarski and Seidenberg on quantifier elimination in a straightforward and accessible way. Using precise logical terminology, this theorem states that the theory of real closed fields admits quantifier elimination. Since we decided to avoid abstract logical formalism, we are limited to the case of the field $\RR$ of real numbers. Note that $\RR$ is the basic example of a real closed field. The details on the Tarski-Seidenberg theorem can be found in \cite{M}, \cite{Rot} or any other textbook on mathematical logic.

The final part of the section is devoted to the second crucial result on quantifier elimination which states that the theory of algebraically closed fields admits quantifier elimination. This result is also proved by Tarski \cite{Ta}. Observe that the assertion holds in particular for the field $\CCC$ of complex numbers. Although the paper does not make any use of quantifier elimination over $\CCC$, we argue that it has a great potential for applications. As an example, we recall the problem of determining whether a completely positive superoperator is irreducible. This problem is studied in \cite{JP1,JP2} (see also \cite{JKP1,JKP2}) and solved completely in \cite{Pa} using techniques of quantifier elimination for $\CCC$. 

First we describe the set $\CF(\RR)$ of all \textit{first-order formulas} over $\RR$. Formulas in $\CF(\RR)$ are built of \textit{atomic formulas}. Atomic formulas in $\CF(\RR)$ are of the form $g_{1}=g_{2}$ or $g_{1}<g_{2}$ where $g_{1}$ and $g_{2}$ are multivariate polynomials over $\RR$. The set $\CF(\RR)$ of all \textit{first-order formulas} is the smallest set satisfying the following conditions:
\begin{enumerate}[\rm(1)]
	\item all atomic formulas belong to $\CF(\RR)$,
	\item if $\varphi,\psi\in\CF(\RR)$, then $(\neg\varphi)\in\CF(\RR)$ and $(\varphi\bullet\psi)\in\CF(\RR)$ where $\bullet\in\{\wedge,\vee,\Ra,\Lra\}$,
	\item if $\varphi\in\CF(\RR)$ and $x$ is a variable, then $(\exists_{x\snull}\varphi),(\forall_{x\snull}\varphi)\in\CF(\RR)$.
\end{enumerate} If $g_{1}$ and $g_{2}$ are multivariate polynomials over $\RR$, then we define $g_{1}\leq g_{2}$ as the logical disjunction $(g_{1}<g_{2})\vee (g_{1}=g_{2})$ and $g_{1}\neq g_{2}$ as the negation $\neg(g_{1}=g_{2})$.

A formula $\varphi\in\CF(\RR)$ is \textit{quantifier-free} if and only if it has no subformula of the form $\exists_{x\snull}\psi$ or $\forall_{x\snull}\psi$ where $\psi\in\CF(\RR)$. Equivalently, the set of all quantifier-free formulas is the smallest set satisfying only the conditions $(1)$ and $(2)$ of the above recursive definition of $\CF(\RR)$. In other words, quantifier-free formulas are boolean combinations of atomic formulas.

Quantifier-free formulas are built of multivariate polynomials over $\RR$. If a formula $\varphi\in\CF(\RR)$ is quantifier-free, then we write $\varphi(x_{1},\dots,x_{n})$ to emphasize that any polynomial that occurs in $\varphi$ belongs to $\RR[x_{1},\dots,x_{n}]$. If $a_{1},\dots,a_{n}$ are concrete real numbers and $\varphi=\varphi(x_{1},\dots,x_{n})$ is quantifier-free, then $\varphi(a_{1},\dots,a_{n})$ is an \textit{evaluation} of $\varphi$ at $a_{1},\dots,a_{n}$. Observe that $\varphi(a_{1},\dots,a_{n})$ is an easily verifiable logical condition (which may be true or false). Therefore quantifier-free formulas can be viewed as \textit{computable conditions}.

\begin{ex} Assume that $f=x^{5}z^{2}+3x^{3}t-7abc^{4}$ and $g=y^{2}+2d^{2}-e^{5}$ are polynomials in $\RR[x,y,z,a,b,c,d,e,t]$. Then the formula $$\varphi(x,y,z,a,b,c,d,e,t)=(\neg(f<g)\wedge(f-g=2x^{2}))\Ra (a+b<c)$$ is a quantifier-free formula in $\CF(\RR)$. If we set $$[x=0,y=1,z=22,a=3,b=1,c=2,d=0,e=2,t=3],$$ then $f(0,1,22,3,1,2,0,2,3)=-42$ and $g(0,1,22,3,1,2,0,2,3)=-31$. Therefore we obtain the evaluation $$\varphi(0,1,22,3,1,2,0,2,3)=(\neg(-42<-31)\wedge(-42+31=0))\Ra (3+1<2)$$ which is easily determined as true. The formula $$(\exists_{x\snull}\forall_{t}(f<g))\Lra(\forall_{a,b,c\snull}f+3=(e-4)^{2})$$ is an example of a general formula in $\CF(\RR)$. \epv
\end{ex}

It is well known that any formula in $\CF(\RR)$ can be written in a \textit{prenex normal form} (this holds for all first-order formulas over a first-order language). This means that a formula $\varphi\in\CF(\RR)$ is equivalent with a formula of the form $Q_{1}Q_{2}\dots Q_{n\snull}\psi$ where any $Q_{i}$ is a quantifier $\exists_{x_{i}}$ or $\forall_{x_{i}}$, $x_{i}$ is some variable and $\psi$ is quantifier-free. 

Assume that $\varphi=Q_{1}Q_{2}\dots Q_{n\snull}\psi(x_{1},\dots,x_{m})$ is written in a prenex normal form. We say that a variable $x_{i}$ is \textit{bound} in $\varphi$ if and only if some $Q_{j}$ is of the form $\exists_{x_{i}}$ or $\forall_{x_{i}}$. We say that $\varphi$ is a \textit{sentence} if and only if all variables $x_{1},\dots,x_{m}$ are bound. We assume for simplicity that if $\varphi$ is a sentence, then it is written in a prenex normal form and if $Q_{i}=\exists_{y}$ or $Q_{i}=\forall_{y}$, then $y\in\{x_{1},\dots,x_{m}\}$. The latter condition means that we do not quantify redundant variables. Observe that, unlike general formulas, sentences are true or false.

Now we are ready to present the aforementioned result of Tarski and Seidenberg on quantifier elimination, restricted to the case of the field $\RR$.

\begin{thm} Assume that $\varphi\in\CF(\RR)$ is a sentence. There exists a quantifier-free formula $\varphi'(y_{1},\dots,y_{m})\in\CF(\RR)$ and real numbers $a_{1},\dots,a_{m}$ such that $\varphi$ is true if and only if the evaluation $\varphi'(a_{1},\dots,a_{m})$ is true. If $\varphi=Q_{1}Q_{2}\dots Q_{n\snull}\psi(x_{1},\dots,x_{n})$, then $a_{1},\dots,a_{m}$ are among coefficients of polynomials which occur in $\psi(x_{1},\dots,x_{n})$. These numbers can be effectively determined, as well as the precise form of $\varphi'(y_{1},\dots,y_{m})$.
\end{thm}

The above theorem has far-reaching consequences for applications of quantifier elimination theory in mathematics, physics or any other field which models its questions within mathematics. Indeed, assume that we are dealing with a \textit{scientific problem} which has the following general form: \begin{center}\textit{Determine whether some mathematical object $\omega$ possesses some property $\pi$.}\end{center} In many cases such problems can be stated as first-order formulas in $\CF(\RR)$. Assume that this is the case, that is, our problem is equivalent with some sentence $\varphi$ in $\CF(\RR)$. First, note that this sentence can be written in a prenex normal form, so we may assume that $\varphi$ is of this form. Then the Tarski-Seidenberg theorem yields we can \textit{compute} a quantifier-free formula $\varphi'(y_{1},\dots,y_{m})$ and some real numbers $a_{1},\dots,a_{m}$ such that $\varphi$ holds if and only if the evaluation $\varphi'(a_{1},\dots,a_{m})$ is true. The condition $\varphi'(a_{1},\dots,a_{m})$ can be easily verified. Consequently, we get a complete solution to the problem we started with.

In this paper we follow the lines of the above scheme. Indeed, assume that the map $\Phi:\MM_{n}(\CCC)\ra\MM_{n}(\CCC)$ is a hermiticity-preserving superoperator. We are interested in determining whether $\Phi$ is positive. For this purpose, we show that there exists a real multivariate polynomial $p_{\Phi}$ such that $\Phi$ is positive if and only if $p_{\Phi}\geq 0$. Obviously, the condition $p_{\Phi}\geq 0$ is a first-order formula in $\CF(\RR)$ and given a concrete $\Phi$, the formula $p_{\Phi}\geq 0$ becomes a sentence. In this way, we are able to apply techniques of quantifier elimination theory for the real closed field $\RR$.

There is one important issue related with quantifier elimination theory. Namely, algorithms which compute the quantifier-free form of a given first-order sentence, or determine its truth value, are rather laborious. In some sense, this is an unsurprising cost of the fact that these procedures can be applied to \textit{any given sentence}. Consequently, we are able to determine whether an \textit{arbitrary} hermiticity-preserving superoperator $\Phi:\MM_{n}(\CCC)\ra\MM_{n}(\CCC)$ is positive or not. Let us stress that in our opinion such a goal cannot be achieved by any other methods. 

We do not exhibit the details on computational complexity of quantifier elimination algorithms. The interested reader is referred to papers \cite{Re1,Re2,Re3} and monographs \cite{Mi,BPR}. We only mention that in many cases these algorithms \textit{determine} whether a given first-order sentence holds or not rather then \textit{compute} its explicit quantifier-free form. Note that the paper \cite{Re1}, which is our basis, takes the same perspective. In fact, these two approaches are close and the former one is usually sufficient for applications.  

The second important result on quantifier elimination (proved also by Tarski) states that the theory of algebraically closed fields admits quantifier elimination. This assertion holds in particular for the field $\CCC$ of complex numbers. Clearly, this fact has similar potential for applications as in the previous case. Moreover, it is much easier than the one for real closed fields, especially in terms of its proof and computational complexity, see \cite{H} and \cite{Mi} for details. Indeed, it is well known that quantifier elimination for algebraically closed fields is equivalent with effective versions of Hilbert's \textit{Nullstellensatz}.

In \cite{Pa} we give a simple and straightforward proof of Tarski's theorem. We base it on the effective version of Hilbert's Nullstellensatz given by Z. Jelonek in \cite{J}. Although similar methods were known, our original motivation was to find computable conditions for irreducibility of completely positive maps. Recall that a completely positive map $\Phi:\MM_{n}(\CCC)\ra\MM_{n}(\CCC)$ is \textit{irreducible} if and only if there is no non-trivial projector $P$ such that $\Phi(P)\leq\lambda P$, for some $\lambda>0$. Alternatively, superoperator $\Phi$ is irreducible if and only if no face of the positive cone in $\MM_{n}(\CCC)$ is invariant under $\Phi$. We refer to \cite{HZ,BZ} for details on irreducible completely positive maps. 

Recall that a superoperator $\Phi$ is completely positive if and only if there are matrices $K_{1},\dots,K_{s}\in\MM_{n}(\CCC)$ (called \textit{Kraus coefficients} of $\Phi$) such that $$\Phi(X)=\sum_{i=1}^{s}K_{i}XK_{i}^{*},$$ for any $X\in\MM_{n}(\CCC)$. The well-known result of D. Farenick proved in \cite{Fa} states that $\Phi$ is irreducible if and only if its Kraus coefficients do not have a non-trivial \textit{common invariant subspace}. This means that if $V$ is a subspace of $\CCC^{n}$ such that $K_{i}V\subseteq V$, for any $i=1,\dots,s$, then $V=0$. We show in \cite{Pa} that this condition can be stated as some first-order sentence $\varphi$ in $\CF(\CCC)$. The set $\CF(\CCC)$ of all  first-order formulas over $\CCC$ is defined similarly as $\CF(\RR)$, but with atomic formulas of the form $g_{1}=g_{2}$ where $g_{1},g_{2}$ are multivariate polynomials over $\CCC$. Applying the reproved version of Tarski's theorem, we compute the quantifier-free formula $\varphi'$ in $\CF(\CCC)$ such that $\varphi$ is equivalent with some evaluation of $\varphi'$. In this way we give a complete solution of the problem studied in \cite{JP1,JP2} and many other papers, see for example \cite{Sh,AI,AGI,GI,Ts}.

\section{Positive maps and real multivariate polynomials}

Throughout, by a \textit{vector space} we mean a finite dimensional vector space $V$ over the field $\CCC$ of complex numbers. We denote by $\mathcal{L}(V)$ the $\mathbb{C}$-vector space of all linear maps $T:V\to V$. Elements of this space are called \textit{operators} on $V$. We are mostly interested in cases $V=\CCC^{n}$, $V=\CCC^{n}\otimes\CCC^{n}$ and $V=\MM_{n}(\CCC)$ where $\MM_{n}(\CCC)$ denotes the space of all $n\times n$ complex matrices. Note that $\CL(\CCC^{n})\cong\MM_{n}(\CCC)$ and hence $$\CL(\CCC^{n}\otimes\CCC^{n})\cong\CL(\CCC^{n})\otimes\CL(\CCC^{n})\cong\MM_{n}(\CCC)\otimes\MM_{n}(\CCC).$$ We often identify the elements of these isomorphic spaces. Elements of $\CL(\MM_{n}(\CCC))$ are called \textit{superoperators}.

This section is devoted to show that if $\Phi\in\CL(\MM_{n}(\CCC))$ is a superoperator that preserves hermiticity, then $\Phi$ is positive if and only if $p_{\Phi}\geq 0$ where $p_{\Phi}$ is some real homogeneous polynomial of degree $4$ in $4n$ variables. This result is only implicitly contained in Chapter 2 of \cite{SkZ}, see also \cite{Ja1,Ja2}. Here we present a rigorous proof of this fact in Theorem 3.2 and the precise form of $p_{\Phi}$ in Theorem 3.5, see also Definition 3.3 and Theorem 3.4. First we introduce some notation and recall basic definitions and facts. We refer to \cite{BZ,HZ,NC,BL} for more details on quantum information theory.

Assume that $n\in\NN$. We denote by $e_1,\dots,e_n$ the elements of the standard $\CCC$-basis of $\CCC^{n}$. These vectors are sometimes considered as $n\times 1$ matrices (that is, as columns). Note that the tensors $e_i\otimes e_j$, for $i,j=1,\dots,n$, form the standard $\CCC$-basis of the vector space $\CCC^n\otimes\CCC^n$. These tensors are denoted by $\epsilon_{ij}$.  

A matrix $A\in\MM_{n}(\CCC)$ is denoted by $[a_{ij}]_{i,j=1,\dots,n}$ or simply by $[a_{ij}]$ when the range of $i,j$ is clear. If $i,j\in\{1,\dots,n\}$, then $E_{ij}$ is the $n\times n$ complex matrix $[e_{kl}]$ such that $e_{kl}\in\{0,1\}$ and $e_{kl}=1$ if and only if $kl=ij$. The matrices $E_{ij}$, for $i,j=1,\dots,n$, form the standard $\CCC$-basis of $\MM_{n}(\CCC)$. 

Recall that $\mathbb{C}^n$ is a Hilbert space with respect to the standard inner product $\langle\cdot\mid\cdot\rangle$ such that for any $x=(x_{1},\dots,x_{n}),y=(y_{1},\dots,y_{n})\in\CCC^{n}$ we have $$\langle x\mid y \rangle=\sum_{i=1}^{n}\ov{x_{i}}y_{i}$$ where $\ov{x_{i}}$ denotes the complex conjugate of $x_{i}$. The norm $\sqrt{\langle x\mid x\rangle}$ of $x\in\CCC^{n}$ is denoted by $\lVert x\rVert$. The space $\CCC^{n}\otimes\CCC^{n}$ is also a Hilbert space with respect to the inner product defined as $$\langle x\otimes y\mid x'\otimes y'\rangle=\langle x\mid x'\rangle\cdot\langle y\mid y'\rangle,$$ for any $x,x',y,y'\in\CCC^{n}$. It is easy to see that $$\langle\sum_{i,j=1}^{n}x_{ij}\epsilon_{ij}\mid\sum_{i,j=1}^{n}y_{ij}\epsilon_{ij}\rangle=\sum_{i,j=1}^{n}\overline{x_{ij}}y_{ij},$$ for any $x_{ij},y_{ij}\in\CCC$ where $i,j=1,\dots,n$.

Assume that $\CH$ is a finite dimensional Hilbert space with the inner product $\langle \cdot\mid\cdot\rangle$. If $T\in\mathcal{L}(\CH)$, then there exists a unique \textit{adjoint operator} $T^*\in\mathcal{L}(\CH)$ such that $\langle Tx\mid y \rangle=\langle x\mid T^*y\rangle$, for any $x,y\in\CH$. Note that if $T\in\MM_{n}(\CCC)$, then $T^*$ is the conjugate transpose of $T$, that is, the matrix \textit{adjoint} to $T$. 

An operator $T\in\CL(\CH)$ is \emph{selfadjoint} (or \emph{hermitian}) if and only if $T=T^{*}$. It is well known that $T$ is selfadjoint if and only if $\langle x\mid Tx\rangle\in\mathbb{R}$, for any $x\in\CH$. A selfadjoint operator $T\in\CL(\CH)$ is \textit{semipositive} if and only if $\langle x\mid Tx\rangle\geq 0$, for any $x\in\CH$. Assume that $\Phi\in\CL(\MM_{n}(\CCC))$ is a superoperator. Then there are $s\geq 1$ and matrices $$A_{1},\dots,A_{s},B_{1},\dots,B_{s}\in\MM_{n}(\CCC)$$ such that $\Phi(X)=\sum_{r=1}^{s}A_{r}XB_{r}$, for any $X\in\MM_{n}(\CCC)$. We say that $\Phi$ \textit{preserves hermiticity} if and only if $\Phi(T)$ is hermitian, whenever $T\in\MM_{n}(\CCC)$ is hermitian. It is well known that $\Phi$ preserves hermiticity if and only if $\Phi$ is of the form $$\Phi(X)=\sum_{r=1}^{s}\alpha_{r}A_{r}XA_{r}^{*}$$ where $\alpha_{r}\in\RR$ is non-zero, for any $r=1,\dots,s$. In case $\alpha_{r}>0$, we say that $\Phi$ is \textit{completely positive}. Note that this is equivalent with assuming that $\alpha_{r}=1$, for any $r=1,\dots,s$. It is easy to see that $\Phi$ preserves hermiticity if and only if $\Phi$ is a difference of two completely positive maps. 

We call $\Phi$ \emph{positive} if and only if $\Phi(T)$ is semipositive, whenever $T\in\MM_{n}(\CCC)$ is semipositive. It is well known that completely positive maps are positive.

Assume that $\Phi\in\CL(\MM_{n}(\CCC))$ is a superoperator that preserves hermiticity. Our aim is to introduce a real homogeneous polynomial $p_{\Phi}$ of degree $4$ in $4n$ variables such that $\Phi$ is positive if and only if $p_{\Phi}\geq 0$. First note that the following crucial theorem holds \cite{Ja1,Ja2}.

\begin{thm} The map $J:\CL(\MM_n(\CCC))\to\MM_n(\CCC)\otimes\MM_n(\CCC)$ defined by the formula $$J(\Phi)=\sum_{i,j=1}^{n}E_{ij}\otimes\Phi(E_{ij}),$$ for any $\Phi\in\CL(\MM_n(\CCC))$, is an isomorphism of Hilbert spaces. Moreover, a superoperator $\Phi$ is positive if and only if $J(\Phi)$ is block positive, for any $\Phi\in\CL(\MM_n(\CCC))$.
\end{thm} 

Recall that an operator $T\in\CL(\CCC^n\otimes\CCC^n)\cong\MM_n(\CCC)\otimes\MM_n(\CCC)$ is \emph{block positive} if and only if $\langle x\otimes y\mid T(x\otimes y)\rangle$ is a non-negative real number, for any $x,y\in\CCC^{n}$. The isomorphism $J$ from the above theorem is known as the \textit{Choi-Jamio{\l}kowski isomorphism}. 

Observe that if $\Phi\in\CL(\MM_n(\CCC))$ preserves hermiticity, then $J(\Phi)$ is selfadjoint. Indeed, it is easy to see that $\Phi(X)^{*}=\Phi(X^{*})$, so we get $$J(\Phi)^{*}=(\sum_{k,l=1}^{n}E_{kl}\otimes\Phi(E_{kl}))^{*}=\sum_{k,l=1}^{n}E_{kl}^{*}\otimes\Phi(E_{kl})^{*}=\sum_{k,l=1}^{n}E_{lk}\otimes\Phi(E_{lk})=J(\Phi).$$ Therefore our aim is to express the block positivity of a selfadjoint operator $T$ as the condition $p_{T}\geq 0$ for some multivariate polynomial $p_{T}$. This is done in the following theorem.

Assume that $T\in\CL(\CCC^{n}\otimes\CCC^{n})$ is a selfadjoint operator and $i,j=1,\dots,n$. Throughout, we denote by $T_{(ij)(kl)}$ the complex numbers such that $$T(\epsilon_{ij})=\sum_{k,l=1}^{n}T_{(ij)(kl)}\epsilon_{kl}.$$ 

\begin{thm} Assume that $T\in\CL(\CCC^{n}\otimes\CCC^{n})$ is a selfadjoint operator. There exists a real multivariate polynomial $p_{T}$ such that $T$ is block positive if and only if $p_{T}\geq 0$. Specifically, $p_{T}$ is a homogeneous polynomial of degree $4$ in $4n$ variables from the set $\{x_{i}^{1},x_{i}^{2},y_{i}^{1},y_{i}^{2}\mid i=1,\dots,n\}$.
\end{thm}

{\bf Proof.} Since $T$ is selfadjoint, we get $\ov{T_{(ij)(kl)}}=T_{(kl)(ij)}$, for any $i,j,k,l=1,\dots,n$. It follows that $T_{(ij)(ij)}\in\RR$, for any $i,j=1,\dots,n$. Assume that $x=(x_{1},\dots,x_{n}),y=(y_{1},\dots,y_{n})\in\CCC^{n}$. We denote by $\iota$ the imaginary unit and fix the following notation: 
\begin{itemize}
	\item $x_{i}=x_{i}^{1}+x_{i}^{2}\iota$ and $y_{i}=y_{i}^{1}+y_{i}^{2}\iota$ where $x_{i}^{1},x_{i}^{2},y_{i}^{1},y_{i}^{2}\in\RR$, for any $i=1,\dots,n$.
	\item $T_{(ij)(kl)}=t_{ijkl}^{1}+t_{ijkl}^{2}\iota$ where $t_{ijkl}^{1},t_{ijkl}^{2}\in\RR$, for any $i,j,k,l=1,\dots,n$ such that $(ij)<(kl)$ ($\leq$ denotes the lexicographic order on $\NN^{2}$).
	\item $T_{(ij)(ij)}=t_{ij}$ where $t_{ij}\in\RR$, for any $i,j=1,\dots,n$.
\end{itemize} Since $$x\otimes y=\sum_{i,j=1}^{n}x_{i}y_{j}\epsilon_{ij},$$ we get the following equalities: $$\langle x\otimes y \mid T(x\otimes y)\rangle=\langle \sum_{i,j=1}^{n}x_{i}y_{j}\epsilon_{ij}\mid\sum_{i,j=1}^{n}x_{i}y_{j}T(\epsilon_{ij})\rangle=$$ $$=\langle\sum_{i,j=1}^{n}x_{i}y_{j}\epsilon_{ij}\mid\sum_{i,j=1}^{n}\sum_{k,l=1}^{n}x_{i}y_{j}T_{(ij)(kl)}\epsilon_{kl}\rangle=\langle\sum_{k,l=1}^{n}x_{k}y_{l}\epsilon_{kl}\mid\sum_{k,l=1}^{n}(\sum_{i,j=1}^{n}x_{i}y_{j}T_{(ij)(kl)})\epsilon_{kl}\rangle=$$ $$=\sum_{k,l=1}^{n}\ov{x_{k}y_{l}}(\sum_{i,j=1}^{n}x_{i}y_{j}T_{(i,j)(k,l)})=\sum_{i,j=1}^{n}\sum_{k,l=1}^{n}T_{(ij)(kl)}\ov{x_{k}y_{l}}x_{i}y_{j}.$$  It is clear that, for any $i,j=1,\dots,n$, the number $$\sigma_{(ij)}:=T_{(ij)(ij)}\ov{x_{i}y_{j}}x_{i}y_{j}=t_{ij}\left\lVert x_i y_j\right\rVert^{2}$$ is a real number. Moreover, for any $i,j,k,l=1,\dots,n$, we have $$\ov{T_{(ij)(kl)}\ov{x_{k}y_{l}}x_{i}y_{j}}=T_{(kl)(ij)}\ov{x_{i}y_{j}}x_{k}y_{l}$$ so if $(ij)<(kl)$, then the number $$\tau_{(ij)(kl)}:=T_{(ij)(kl)}\ov{x_{k}y_{l}}x_{i}y_{j}+T_{(kl)(ij)}\ov{x_{i}y_{j}}x_{k}y_{l}=2Re(T_{(ij)(kl)}\ov{x_{k}y_{l}}x_{i}y_{j})$$ is also real. In fact, we may view both $\sigma_{(ij)}$ and 
$\tau_{(ij)(kl)}$ as real homogeneous polynomials. Indeed, observe that, for any $r,s=1,\dots,n$, we have $$x_{r}y_{s}=(x_{r}^{1}y_{s}^{1}-x_{r}^{2}y_{s}^{2})+(x_{r}^{1}y_{s}^{2}+x_{r}^{2}y_{s}^{1})\iota.$$ We set $\alpha_{rs}=x_{r}^{1}y_{s}^{1}-x_{r}^{2}y_{s}^{2}$, $\beta_{rs}=x_{r}^{1}y_{s}^{2}+x_{r}^{2}y_{s}^{1}$ and after some standard calculations we obtain that $\sigma_{(ij)}=t_{ij}(\alpha_{ij}^{2}+\beta_{ij}^{2})$,
for any $i,j=1,\dots,n$, and $$\tau_{(ij)(kl)}=2t_{ijkl}^{1}(\alpha_{kl}\alpha_{ij}+\beta_{kl}\beta_{ij})-2t_{ijkl}^{2}(\alpha_{kl}\beta_{ij}-\beta_{kl}\alpha_{ij}),$$ for any $(ij)<(kl)$. This means that $$\sigma_{(ij)},\tau_{(ij)(kl)}\in\RR[x_{i}^{1},x_{i}^{2},x_{j}^{1},x_{j}^{2},y_{k}^{1},y_{k}^{2},y_{l}^{1},y_{l}^{2}]$$ are homogeneous polynomials of degree $4$. Define $I$ to be the set of all $(ij)(kl)$ such that $i,j,k,l=1,\dots,n$ and $(ij)<(kl)$. Finally, set $$p_{T}:=\sum_{i,j=1}^{n}\sigma_{(ij)}+\sum_{(ij)(kl)\in I}\tau_{(ij)(kl)}$$ which is a real homogeneous polynomial of degree $4$ in $4n$ variables from the set $\{x_{i}^{1},x_{i}^{2},y_{i}^{1},y_{i}^{2}\mid i=1,\dots,n\}$. The above arguments show that $T$ is block-positive if and only if $p_{T}\geq 0$. \epv

Theorem 3.2 suggests the following useful definition.

\begin{df} Assume that $T\in\CL(\CCC^{n}\otimes\CCC^{n})$ is a selfadjoint operator such that 
\begin{itemize}
	\item $T_{(ij)(kl)}=t_{ijkl}^{1}+t_{ijkl}^{2}\iota$ where $t_{ijkl}^{1},t_{ijkl}^{2}\in\RR$, for any $i,j,k,l=1,\dots,n$ such that $(ij)<(kl)$,
	\item $T_{(ij)(ij)}=t_{ij}$ where $t_{ij}\in\RR$, for any $i,j=1,\dots,n$.
\end{itemize} Moreover, assume that
\begin{itemize}
	\item $\CV=\{x_{i}^{1},x_{i}^{2},y_{i}^{1},y_{i}^{2}\mid i=1,\dots,n\}$ and $x_{i}=x_{i}^{1}+x_{i}^{2}\iota$, $y_{i}=y_{i}^{1}+y_{i}^{2}\iota$, for any $i=1,\dots,n$,
	\item $\alpha_{rs}=x_{r}^{1}y_{s}^{1}-x_{r}^{2}y_{s}^{2}$, $\beta_{rs}=x_{r}^{1}y_{s}^{2}+x_{r}^{2}y_{s}^{1}$, for any $r,s=1,\dots,n$,
	\item $I$ is the set of all $(ij)(kl)$ such that $i,j,k,l=1,\dots,n$ and $(ij)<(kl)$.
\end{itemize} The \textit{positivity polynomial} for $T$ is a polynomial $p_{T}\in\RR[\CV]$ such that $$p_{T}:=\sum_{i,j=1}^{n}\sum_{k,l=1}^{n}T_{(ij)(kl)}\ov{x_{k}y_{l}}x_{i}y_{j}=\sum_{i,j=1}^{n}\sigma^T_{(ij)}+\sum_{(ij)(kl)\in I}\tau_{(ij)(kl)}^{T}$$ where $$\tau_{(ij)(kl)}^{T}:=2t_{ijkl}^{1}(\alpha_{kl}\alpha_{ij}+\beta_{kl}\beta_{ij})-2t_{ijkl}^{2}(\alpha_{kl}\beta_{ij}-\beta_{kl}\alpha_{ij})$$ and $\sigma^T_{(ij)}:=t_{ij}(\alpha^2_{ij}+\beta_{ij}^2)$. \epv

\end{df} It is convenient to formulate Theorem 3.2 in the following way.

\begin{thm} A selfadjoint operator $T\in\CL(\CCC^{n}\otimes\CCC^{n})$ is block positive if and only if its positivity polynomial $p_{T}$ satisfies the condition $p_{T}\geq 0$.
\end{thm} 

{\bf Proof.} The assertion follows from the proof of Theorem 3.2 and the definition of positivity polynomial, see Definition 3.3. \epv


Assume that $\Phi\in\CL(\MM_{n}(\CCC))$ is a superoperator which preserves hermiticity. Since $J(\Phi)$ is selfadjoint, it follows from Theorem 3.1 and Theorem 3.4 that $\Phi$ is positive if and only if the positivity polynomial $p_{\Phi}:=p_{J(\Phi)}$ satisfies $p_{\Phi}\geq 0$. This fact is stated in the assertion $(1)$ of the following theorem. The assertion $(2)$ is devoted to present the exact form of the polynomial $p_{\Phi}$.

\begin{thm} Assume that $\Phi\in\CL(\MM_{n}(\CCC))$ is a hermiticity-preserving superoperator such that $\Phi(X)=\sum_{r=1}^{s}\alpha_{r}A_{r}XA_{r}^{*}$ and $A_{r}=[a_{ij}^{r}]$, for any $r=1,\dots,s$.
\begin{enumerate}[\rm(1)]
	\item Superoperator $\Phi$ is positive if and only if the positivity polynomial $p_{\Phi}:=p_{J(\Phi)}$ satisfies $p_{\Phi}\geq 0$.
	\item We have $$J(\Phi)_{(ij)(kl)}=\sum_{r=1}^{s}\alpha_{r}a_{lk}^{r}\ov{a_{ji}^{r}},$$ for any $i,j,k,l=1,\dots,n$. Moreover, we have $$p_{\Phi}=\sum_{i,j=1}^{n}\sum_{k,l=1}^{n}\sum_{r=1}^{s}\alpha_{r}a_{lk}^{r}\ov{a_{ji}^{r}}\ov{x_{k}y_{l}}x_{i}y_{j}=\sum_{r=1}^{s}\alpha_{r}\left\lVert[x_{1},\dots,x_{n}]\cdot A_{r}^{*}\cdot[y_{1},\dots,y_{n}]^{tr}\right\rVert^{2}$$ where the dot represents the matrix multiplication.
\end{enumerate}
\end{thm}

{\bf Proof.} (1) The operator $J(\Phi)$ is selfadjoint, so $J(\Phi)$ is block positive if and only if the polynomial $p_{\Phi}=p_{J(\Phi)}$ satisfies $p_{\Phi}\geq 0$, see Theorem 3.4. Theorem 3.1 yields $\Phi$ is positive if and only if $J(\Phi)$ is block positive, so the assertion follows.

(2) Observe that $$J(\Phi)\epsilon_{ij}=\sum_{k,l=1}^{n}(E_{kl}e_{i})\otimes(\Phi(E_{kl})e_{j})=\sum_{k=1}^{n}e_{k}\otimes(\Phi(E_{ki})e_{j})$$ and thus straightforward calculations yield $$J(\Phi)_{(ij)(kl)}=e_{l}^{tr}\Phi(E_{ki})e_{j}=\sum_{r=1}^{s}\alpha_{r}a_{lk}^{r}\ov{a_{ji}^{r}}.$$ Note that $J(\Phi)_{(ij)(kl)}=J(\Phi)_{(kl)(ij)}$, for any $i,j,k,l\in 1,\dots,n$, which also shows that $J(\Phi)$ is selfadjoint. In consequence, we get $$p_{\Phi}=p_{J(\Phi)}=\sum_{i,j=1}^{n}\sum_{k,l=1}^{n}J(\Phi)_{(ij)(kl)}\ov{x_{k}y_{l}}x_{i}y_{j}=$$$$=\sum_{i,j=1}^{n}\sum_{k,l=1}^{n}\sum_{r=1}^{s}\alpha_{r}a_{lk}^{r}\ov{a_{ji}^{r}}\ov{x_{k}y_{l}}x_{i}y_{j}.$$ Therefore the following equalities hold: $$p_{\Phi}=\sum_{r=1}^{s}\alpha_{r}(\sum_{i,j=1}^{n}\sum_{k,l=1}^{n}a_{lk}^{r}\ov{a_{ji}^{r}}\ov{x_{k}y_{l}}x_{i}y_{j})=\sum_{r=1}^{s}\alpha_{r}((\sum_{i,j=1}^{n}\ov{a_{ji}^{r}}x_{i}y_{j})(\sum_{k,l=1}^{n}a_{lk}^{r}\ov{x_{k}y_{l}}))=$$ $$=\sum_{r=1}^{s}\alpha_{r}([x_{1},\dots,x_{n}]\cdot A_{r}^{*}\cdot[y_{1},\dots,y_{n}]^{tr})([\ov{y_{1}},\dots,\ov{y_{n}}]\cdot A_{r}\cdot[\ov{x_{1}},\dots,\ov{x_{n}}]^{tr})$$ $$=\sum_{r=1}^{s}\alpha_{r}\left\lVert[x_{1},\dots,x_{n}]\cdot A_{r}^{*}\cdot[y_{1},\dots,y_{n}]^{tr}\right\rVert^{2}$$ and the assertion follows. \epv

The assertion $(2)$ of the above theorem gives an interesting description of the positivity polynomial $p_{\Phi}$ as the sum of some non-positive or non-negative real multivariate polynomials, depending on signs of the numbers $\alpha_{1},\dots,\alpha_{s}$. As a consequence, we get an alternative proof of the fact that completely positive maps are positive. Indeed, if $\Phi$ is completely positive, then $\alpha_{1},\dots,\alpha_{s}>0$, so in this case it is obvious that $p_{\Phi}\geq 0$. This shows that the above description of $p_{\Phi}$ is useful. 

\begin{remark} Assume that $\Phi\in\CL(\MM_{n}(\CCC))$ is a hermiticity-preserving superoperator such that $\Phi(X)=\sum_{r=1}^{s}\alpha_{r}A_{r}XA_{r}^{*}$. If $\Phi$ is a fixed operator (in the sense that concrete $\alpha_{r}\in\RR$ and $A_{r}\in\MM_{n}(\CCC)$ are given, for any $r=1,\dots,s$), its positivity polynomial $p_{\Phi}$ may be calculated using any computer algebra system. We recall that $$p_{\Phi}=\sum_{r=1}^{s}\alpha_{r}\left\lVert[x_{1},\dots,x_{n}]\cdot A_{r}^{*}\cdot[y_{1},\dots,y_{n}]^{tr}\right\rVert^{2}$$ by the equality presented in condition (2) of Theorem 3.5. Hence $p_{\Phi}$ may be calculated in three different, but equivalent ways, see also Definition 3.3 and the proof of Theorem 3.2. \epv
\end{remark}



\section{Generalized Sturm's theorem and some applications}

In this section we recall the renowned \textit{generalized Sturm's theorem} which is also known as the \textit{Sturm-Tarski theorem}. This theorem allows to calculate the number of distinct real roots of a polynomial $p$ satisfying some additional conditions. We apply this result in a procedure that determines the validity of the sentence $$\exists_{x\snull}(p(x)>0\wedge q(x)>0)$$ where $p,q\in\RR[x]$. This is given in Algorithm 4.4 which is applied in Section 5. 

In some parts, this section is based on \cite{KB}. The details on generalized Sturm's theorem, as well as its proof, can be found in \cite{BPR} or \cite{XY}. Crucial applications of this theorem in quantifier elimination for the theory of real numbers are given in \cite{Re1,Re2,Re3}. 


Assume that $n\in\NN$. A tuple $(h_{0},h_{1},\hdots,h_{n})\in\RR[x]^{n}$ of non-zero polynomials is a \textit{Sturm sequence} if and only if the following conditions hold:
\begin{enumerate}[\rm(1)]
	\item The polynomial $h_{n}$ does not have real roots.
	\item If $h_{i}(x)=0$, for some $x\in\RR$, then $h_{i-1}(x)h_{i+1}(x)<0$, for any $i=1,\dots,n-1$.
\end{enumerate}

There is a canonical construction of a Sturm sequence associated with two non-zero polynomials $p,q\in\RR[x]$. Up to the sign, its elements are polynomials that occur as remainders in the Euclid's algorithm for determining the greatest common divisor of $p$ and $q$. We recall this recursive construction below.

Assume that $p,q\in\RR[x]$ are non-zero polynomials. First we construct the \textit{canonical sequence} for $p$ and $q$ (which is also called the \textit{generalized Sturm sequence}). Set $h_{0}=p$ and $h_{1}=q$. Assume that $n\geq 1$ and $h_{0},h_{1},\hdots,h_{n}$ are defined. If $h_{n}\mid h_{n-1}$, then the canonical sequence is the sequence $(h_{0},h_{1},\hdots,h_{n})$. Otherwise, we set $h_{n+1}=-\rem_{h_{n}}(h_{n-1})$ where $\rem_{h_{n}}(h_{n-1})$ is the remainder of division of polynomial $h_{n-1}$ by $h_{n}$. 

Assume that $(h_{0},h_{1},\hdots,h_{n})$ is the canonical sequence for $p$ and $q$. It follows from the Euclid's algorithm that $h_{n}$ is, up to the sign, the greatest common divisor of $p$ and $q$ and thus $h_{n}\mid h_{i}$, for any $i=0,\hdots,n$. It is well known that the sequence $$\left(\frac{h_{0}}{h_{n}},\frac{h_{1}}{h_{n}},\hdots,\frac{h_{n}}{h_{n}}=1\right)$$ is a Sturm sequence, see for example \cite{XY}. We call this sequence the \textit{canonical Sturm sequence} for $p$ and $q$. There are other examples of Sturm sequences as well. For example, if a polynomial $h$ with $\deg(p)=d$ has no multiple real roots, then the sequence $$(h,h',h'',h^{(3)},\hdots,h^{(d)})$$ of all consecutive derivatives of $h$ is a Sturm sequence, see \cite{BPR}.

Assume that $h\in\RR[x]$ and $a\in\{-\infty,\infty\}$. If $\lim_{x\ra a}h(x)=\infty$, then we set $\sigma_{a}(h)=+$. Otherwise, we set $\sigma_{a}(h)=-$. If $h_{1},\hdots,h_{s}\in\RR[x]$, then we define $$\sigma_{a}(h_{1},\hdots,h_{s})=(\sigma_{a}(h_{1}),\hdots,\sigma_{a}(h_{s}))\in\{+,-\}^{s}.$$ Assume that $(\alpha_{1},\hdots,\alpha_{s})\in\{+,-\}^{s}$. A subsequence $(\alpha_{i},\alpha_{i+1})$ of $(\alpha_{1},\hdots,\alpha_{s})$ is a \textit{sign change} if and only if $(\alpha_{i},\alpha_{i+1})\in\{(-,+),(+,-)\}$. We denote by $\lambda(\alpha_{1},\hdots,\alpha_{s})$ the number of all sign changes in $(\alpha_{1},\hdots,\alpha_{s})$.

Assume that $p,q\in\RR[x]$ are non-zero polynomials and let $(h_{0},h_{1},\hdots,h_{n})$ be the canonical Sturm sequence for $p$ and $q$. We define $\nu(p,q)$ as the number $$\lambda(\sigma_{-\infty}(h_{0},\hdots,h_{n}))-\lambda(\sigma_{\infty}(h_{0},\hdots,h_{n})).$$ 

If $X$ is a finite set, then $|X|$ denotes the number of elements of $X$. Assume that $f,g\in\RR[x]$ are non-zero polynomials. We define $N(f,g)$ as the number $$|\{x\in\RR\mid f(x)=0\wedge g(x)>0\}|-|\{x\in\RR\mid f(x)=0\wedge g(x)<0\}|.$$ Observe that if $g$ is a polynomial such that $g>0$, then $N(f,g)$ is the number of distinct real roots of polynomial $f$.

The following theorem is known as the \textit{generalized Sturm's theorem} or the \textit{Sturm-Tarski theorem}.

\begin{thm} Assume that $f,g\in\RR[x]$ are non-zero polynomials. Then we have $$\nu(f,f'g)=N(f,g).$$ In particular, the number $N(f,g)$ can be computed. \epv
\end{thm}

Observe that if $g=1$ is a constant polynomial, then the above theorem yields that the number $N(f,1)$ of all distinct real roots of $f$ can be computed as $\nu(f,f')$. We note that more general versions of the above theorem are known, see for instance \cite{XY} or \cite{BPR}. Here we apply only the above special version. 

Generalized Sturm's theorem is an important tool in quantifier elimination for the theory of real numbers. Indeed, it allows to eliminate quantifiers from formulas of the form $Q_{1}\dots Q_{n\snull}\varphi$ where $Q_{1},\dots, Q_{n}\in\{\exists,\forall\}$ and $\varphi$ is a quantifier-free formula involving only univariate polynomials. We refer the reader to \cite{Re3} for details on these matters. 

In our restricted setting we are interested in deciding whether there exists $x\in\RR$ such that $p(x)>0$ and $q(x)>0$, where $p,q\in\RR[x]$. This is related with counting the number of elements of the set $$\CS(f,p,q):=\{x\in\RR\mid f(x)=0\wedge p(x)>0\wedge q(x)>0\}$$ where $f,p,q\in\RR[x]$ are non-zero polynomials. The proposition stated below shows how this can be done. Recall that the sign function $\sgn:\RR\ra\{-1,0,1\}$ is defined as follows: $\sgn(r)=-1$, if $r<0$, $\sgn(r)=0$, if $r=0$ and $\sgn(r)=1$, if $r>0$. Note that $\sgn(r_{1}r_{2})=\sgn(r_{1})\sgn(r_{2})$, for any $r_{1},r_{2}\in\RR$. Moreover, denoting by $\CZ(f)$ the set of all distinct real roots of $f$, it is easy to see that $$N(f,g)=\sum_{x\in\CZ(f)}\sgn(g(x)).$$ 

\begin{prop} Assume that $p,q,f\in\RR[x]$ are non-zero polynomials. Then we have the equality $$|\CS(f,p,q)|=\frac{1}{4}(N(f,p^{2}q^{2})+N(f,p^{2}q)+N(f,pq^{2})+N(f,pq)).$$ In particular, the number $|\CS(f,p,q)|$ can be computed.
\end{prop}

{\bf Proof.} Assume that $l(x)=\frac{1}{2}(x^{2}+x)$. Then $l(1)=1$ and $l(x)=0$ if and only if $x=0$ or $x=-1$. Therefore, if $h\in\RR[x]$, then $l(\sgn(h(x)))\neq 0$ if and only if $h(x)>0$ and in this case we have $l(\sgn(h(x)))=1$. It follows that, for $p,q,f\in\RR[x]$, we obtain the equality $$|\CS(f,p,q)|=\sum_{x\in\CZ(f)}l(\sgn(p(x)))l(\sgn(q(x)))=$$ $$=\sum_{x\in\CZ(f)}\frac{1}{4}(\sgn(p(x))^{2}+\sgn(p(x)))(\sgn(q(x))^{2}+\sgn(q(x))).$$ Moreover, since $\sgn:\RR\ra\{-1,0,1\}$ preserves multiplication, we get $$|\CS(f,p,q)|=\frac{1}{4}(\sum_{x\in\CZ(f)}\sgn(p^{2}(x)q^{2}(x))+\sum_{x\in\CZ(f)}\sgn(p^{2}(x)q(x))+$$ $$+\sum_{x\in\CZ(f)}\sgn(p(x)q^{2}(x))+\sum_{x\in\CZ(f)}\sgn(p(x)q(x)))$$ which shows the first part of the assertion. The second one follows directly from Theorem 4.1. \epv

The argumentation given in the proof of Proposition 4.2 can be generalized to determine the number of $x\in\RR$ such that $f(x)=0$, $r_{1}(x)=\hdots=r_{m}(x)=0$ and $q_{1}(x)>0\wedge\hdots\wedge q_{n}(x)>0$ where $f,r_{1},\hdots,r_{m},q_{1},\hdots,q_{n}\in\RR[x]$ are fixed non-zero polynomials. This is done in Section 5 of \cite{KB}.

We denote by $\ov{\RR}$ the set $\RR\cup\{-\infty,\infty\}$ and we assume that $-\infty<a<\infty$, for any $a\in\RR$.

\begin{cor} Let $p,q\in\RR[x]$ be non-zero polynomials such that the condition $$(*)\quad(\lim_{x\ra\infty}p(x)=\lim_{x\ra\infty}q(x)=\infty)\vee(\lim_{x\ra-\infty}p(x)=\lim_{x\ra-\infty}q(x)=\infty)$$ does not hold. In this case the formula $\exists_{x\snull}(p(x)>0\wedge q(x)>0)$ is equivalent with the formula $$\exists_{x\snull}(p(x)>0\wedge q(x)>0\wedge (pq)'(x)=0).$$ Therefore these formulas are equivalent with the condition $|\CS((pq)',p,q)|\neq 0$.
\end{cor}

{\bf Proof.} It suffices to prove that the condition $\exists_{x\snull}(p(x)>0\wedge q(x)>0)$ implies that $\exists_{x\snull}(p(x)>0\wedge q(x)>0\wedge (pq)'(x)=0)$ (the other implication is obvious). 

Assume that there is $x\in\RR$ such that $p(x)>0$ and $q(x)>0$. Moreover, assume that $r\in\{p,q\}$ and $I_{r}=(a_{r},b_{r})$ is the largest open interval on the real line $\RR$ such that $x\in I_{r}$ and $r(t)>0$, for any $t\in I_{r}$. It is easy to see that if $a_{r}\neq-\infty$ ($b_{r}\neq\infty$, respectively), then $a_{r}$ ($b_{r}$, respectively) is a root of $r$. Note that $x\in I_{p}\cap I_{q}$, so $I_{p}\cap I_{q}$ is an interval. Assume that $I_{p}\cap I_{q}=(c,d)$ where $c,d\in\ov{\RR}$. Then $c\neq-\infty$ and $d\neq\infty$, because the condition $(*)$ does not hold. Hence we get that $c,d\in\RR$ are some different roots of the polynomial $pq$ and thus there exists $t\in(c,d)$ such that $(pq)'(t)=0$. Since $p$ and $q$ are positive on $(c,d)$, we get that $p(t)>0\wedge q(t)>0\wedge (pq)'(t)=0$. The second assertion is a consequence of the definition of $\CS(f,p,q)$. \epv

We finish this section with the following algorithm that is applied in Section 5. 

\begin{alg} Input: two non-zero polynomials $p,q\in\RR[x]$. Output: \textit{yes} or \textit{no} depending on the validity of the formula $\exists_{x\snull}(p(x)>0\wedge q(x)>0)$.
\begin{enumerate}[\rm(1)]
	\item Determine whether the condition $$(\lim_{x\ra\infty}p(x)=\lim_{x\ra\infty}q(x)=\infty)\vee(\lim_{x\ra-\infty}p(x)=\lim_{x\ra-\infty}q(x)=\infty)$$ holds by looking at degrees and leading coefficients of polynomials $p,q$. If it does, then the output is \textit{yes}. Otherwise go to $(2)$.
	\item Determine whether $|\CS((pq)',p,q)|\neq 0$. If this holds, the output is \textit{yes}. Otherwise, the output is \textit{no}.
\end{enumerate}
\end{alg}

{\bf Proof.} The correctness of the algorithm follows directly from Corollary 4.3. \epv

\begin{remark}
If $q$ is any polynomial such that $q>0$, then the formula $\exists_{x\snull}p(x)>0$ is equivalent with $\exists_{x\snull}(p(x)>0\wedge q(x)>0)$. Hence Algorithm 4.4 can be applied to determine whether the formula $\exists_{x\snull}p(x)>0$ holds, for any given polynomial $p\in\RR[x]$. This formula is the negation of $\forall_{x\snull}-p(x)\geq 0$, so Algorithm 4.4 can be applied to determine whether $r\geq 0$, for any given polynomial $r\in\RR[x]$. \epv
\end{remark}

\section{Positivity of some real multivariate polynomials}

The series of papers \cite{Re1,Re2,Re3} by J. Renegar is devoted to present a quantifier elimination technique for the theory of real numbers. In particular, the main result of \cite{Re1} is a decision method for the \textit{existential theory of $\RR$}. In this section we adjust these general techniques to the case that appears in the context of hermiticity-preserving superoperators. Specifically, we present an algorithm for determining whether $g\geq 0$, if $g\in\RR[x_{1},\hdots,x_{n}]$ is a homogeneous polynomial of an even degree, see Algorithm 5.6. This algorithm is based on few constructions, see Construction 5.1, 5.3 and 5.4, which are special cases of the ones presented in \cite{Re1} (with some changes in notation). 

We note that restricting to the case of homogeneous polynomials of even degrees is consistent with results of Section 3, because positivity polynomials are homogeneous of degree $4$, see Theorem 3.2 and Definition 3.3. In the case of such polynomials, constructions from \cite{Re1} get slightly simpler. 

The algorithm for determining positivity of a given hermiticity-preserving superoperator is a direct consequence of Algorithm 5.6 and Theorem 3.5. This is presented in Algorithm 5.7. 

We start with introducing some notation related with multivariate polynomials over arbitrary commutative rings. 

Assume that $n\in\NN$ and $n\geq 1$. We view the set $\NN^{n}$ as a poset with respect to the \textit{lexicographical order}. Recall that this ordering is linear. If $\alpha=(a_{1},\dots,a_{n})\in\NN^{n}$, then we set $|\alpha|=a_{1}+\dots +a_{n}$. 

The tuple $(x_{1},\hdots,x_{n})$ is denoted by $\ov{x}$ and $\ov{x}^{\alpha}$ denotes the monomial $x_{1}^{a_{1}}\dots x_{n}^{a_{n}}$. The \textit{degree} of this monomial is the number $|\alpha|$. The set of all monomials is ordered lexicographically via the natural identification $\alpha\leftrightarrow \ov{x}^{\alpha}$, for any $\alpha\in\NN^{n}$.

Assume that $R$ is a commutative ring with a unit. A polynomial $g\in R[\ov{x}]$ is denoted as $g=\sum_{\alpha}a_{\alpha}\ov{x}^{\alpha}$ where $a_{\alpha}\in R$ and $a_{\alpha}\neq 0$ only for finitely many $\alpha\in\NN^{n}$. The \textit{degree} of $g=\sum_{\alpha}a_{\alpha}\ov{x}^{\alpha}$ is the maximal number of the set $\{|\alpha|\in\NN\mid a_{\alpha}\neq 0\}$. The degree of $g$ is denoted by $\deg(g)$. If $a_{\alpha}\neq 0$ only in the case $|\alpha|=d$, then we say that $g$ is \textit{d-homogeneous}.

Assume that $g\in R[\ov{x}]$, $g=\sum_{\alpha}a_{\alpha}\ov{x}^{\alpha}$ and $\deg(g)=d$. A \textit{d-homogenization} of $g$ is a polynomial $g_{hom}\in R[\ov{x},x_{n+1}]$ such that $g^{hom}=\sum_{\alpha}a_{\alpha}\ov{x}^{\alpha}x_{n+1}^{d-|\alpha|}$. It is easy to see that $g^{hom}$ is a $d$-homogeneous polynomial.

Assume that $g\in\RR[x_{1},\hdots,x_{n}]$ is a homogeneous polynomial of an even degree. The algorithm we shall present relies on a construction of some set $\CR_{g}$ of polynomials in variables $u_{1},\dots,u_{n},u_{n+1}$. This is the most sophisticated part of the procedure. Further steps are shorter and they are based on the following construction and theorem from \cite{Re1}.

\begin{cons} Assume that $g\in\RR[x_{1},\hdots,x_{n}]$ is a homogeneous polynomial, $d=\deg(g)$ is an even number and the set $\CR_{g}\subseteq\RR[u_{1},\dots,u_{n},u_{n+1}]$ is given. Define $J(n,d)=\{0,\dots,nd^{2n}\}$ and $$\CB(n,d)=\{(i^{n-1},i^{n-2},\dots,i,1,0)\in\NN^{n+1}\mid i=0,\dots,nd^{2n}\}.$$ Assume that $j\in J(n,d)$, $\beta\in\CB(n,d)$ and $r\in\CR_{g}$ are fixed. We define some univariate polynomials $r_{1},\dots,r_{n+1}\in\RR[t]$ and $g_{j,\beta,r,+}^{un},g_{j,\beta,r,-}^{un}\in\RR[t]$ as follows: $$r_{i}(t)=(\frac{\partial r}{\partial u_{i}})(\beta+te_{n+1}),$$ for any $i=1,\dots,n+1$ and $$g_{j,\beta,r,+}^{un}(t)=g(r_{1}^{(j)}(t),r_{2}^{(j)}(t),\dots,r_{n}^{(j)}(t)),$$ $$g_{j,\beta,r,-}^{un}(t)=g(-r_{1}^{(j)}(t),-r_{2}^{(j)}(t),\dots,-r_{n}^{(j)}(t))$$ where $r_{i}^{(j)}$ denotes the $j$-th derivative of $r_{i}$. \epv
\end{cons}

\begin{thm} Assume that $g\in\RR[x_{1},\hdots,x_{n}]$ is a homogeneous polynomial and $d=\deg(g)$ is an even number. Then $g\geq 0$ if and only if for any $j\in J(n,d)$, $\beta\in\CB(n,d)$ and $r\in\CR_{g}$ the following condition $$(*)\quad(\exists_{t}-g_{j,\beta,r,+}^{un}(t)>0\wedge r_{n+1}(t)>0)\vee(\exists_{t}-g_{j,\beta,r,-}^{un}(t)>0\wedge -r_{n+1}(t)>0)$$ does not hold.
\end{thm}

{\bf Proof.} The assertion is a direct consequence of the main results of \cite{Re1}, see especially Sections 3 and 4. \epv

It is important to emphasize that the condition $(*)$ from Theorem 5.2 can be verified by Algorithm 4.4 thanks to the fact that polynomials $g_{j,\beta,r,+}^{un},g_{j,\beta,r,-}^{un}$ and $r_{n+1}$ are univariate.

Our goal is to construct the aforementioned $\CR_{g}$. The construction of this set is related with another one which produces some set $\mathcal{R}(g_{1},\dots,g_{n})\subseteq\RR[u_{1},\dots,u_{n},u_{n+1}]$ where $g_{1},\dots,g_{n}\in R[x_{1},\dots,x_{n}]$ are given polynomials. That initial construction is recalled below, following the lines of Section 2.2 of \cite{Re1}. 

\begin{cons}
Assume that $R=\RR[\delta,\gamma]$ and $g_{1},\dots,g_{n}\in R[x_{1},\dots,x_{n}]$ are polynomials of degree at most $d$ such that $\deg(g_{i})=d$, for some $i=1,\dots,n$. Assume that $$g_{1}^{hom},\dots,g_{n}^{hom}\in R[x_{1},\dots,x_{n},x_{n+1}]$$ are $d$-homogenizations of polynomials $g_{1},\dots,g_{n}$, respectively. Here we assume that $\ov{x}$ is the tuple $(x_{1},\dots,x_{n+1})$. Let $$\CS=\{\alpha\in\NN^{n+1}\mid |\alpha|=n(d-1)+1\}\textnormal{ and }\TT=\{\ov{x}^{\alpha}\mid\alpha\in\CS\}.$$ Assume that $\alpha=(a_{1},\dots,a_{n+1})\in\CS$. We define some multivariate polynomial $t_{\alpha}$ in the following way:
\begin{enumerate}[\rm(1)]
	\item Assume that there exists $j\in\{1,\dots,n\}$ such that $d\leq a_{j}$ and let $i$ be the least such a number. Then $t_{\alpha}\in R[x_{1},\dots,x_{n+1}]$ and $$t_{\alpha}=x_{1}^{a_{1}}\dots x_{i}^{a_{i}-d}\dots x_{n+1}^{a_{n+1}}g_{i}^{hom}.$$ 
	\item Assume that $a_{j}\leq d-1$, for any $j\in\{1,\dots,n\}$. Note that in this case we have $a_{1}+\dots +a_{n}\leq n(d-1)$ and since $|\alpha|=n(d-1)+1$, we get $a_{n+1}\geq 1$. Then $t_{\alpha}\in R[x_{1},\dots,x_{n+1},u_{1},\dots,u_{n+1}]$ and $$t_{\alpha}=x_{1}^{a_{1}}\dots x_{n+1}^{a_{n+1}-1}(\sum_{i=1}^{n+1}u_{i}x_{i})=\sum_{i=1}^{n+1}x_{1}^{a_{1}}\dots x_{i}^{a_{i}+1}\dots x_{n+1}^{a_{n+1}-1}u_{i}.$$
\end{enumerate} 

It is easy to see that in both cases $(1)$ and $(2)$ the polynomial $t_{\alpha}$ is a linear combination of elements of $\TT$ with scalars belonging to the set $$R\cup\{u_{1},\dots,u_{n+1}\}=\RR[\delta,\gamma]\cup\{u_{1},\dots,u_{n+1}\}\subseteq\RR[u_{1},\dots,u_{n+1},\delta,\gamma].$$ More precisely, in $(1)$ these scalars are elements of $R=\RR[\delta,\gamma]$ and in $(2)$ they belong to the set $\{u_{1},\dots,u_{n+1}\}$. Summing up, we conclude that for any $\alpha\in\CS$ we have the presentation $$t_{\alpha}=\sum_{\beta\in\CS}b_{\beta}\ov{x}^{\beta}$$ where $b_{\beta}\in\RR[u_{1},\dots,u_{n+1},\delta,\gamma]$. Now, we define some function $$M:\CS\times\CS\ra\RR[u_{1},\dots,u_{n+1},\delta,\gamma]$$ in the following way: if $\alpha\in\CS$ and $t_{\alpha}=\sum_{\beta\in\CS}b_{\beta}\ov{x}^{\beta}$, then we set $M(\alpha,\beta)=b_{\beta}$. Since the set $\CS$ is linearly ordered by the lexicographical order, we may view $M$ as a $|\CS|\times|\CS|$ matrix. Observe that $\det(M)\in\RR[u_{1},\dots,u_{n+1},\delta,\gamma]=\RR[u_{1},\dots,u_{n+1}][\delta,\gamma]$ and assume that $$\det(M)=\sum_{i,j}g_{ij}\delta^{i}\gamma^{j}$$ where $g_{ij}$ are some polynomials in $\RR[u_{1},\dots,u_{n+1}]$. We define the set $\mathcal{R}(g_{1},\dots,g_{n})$ as the set of all polynomials $g_{ij}$. \epv
\end{cons}

We are prepared to finish the construction of the set $\CR_{g}$. We introduce the following notation: if $h$ is a polynomial in $n+m$ variables $x_{1},\dots,x_{1},y_{1},\dots,y_{m}$ and $\ov{x}$ denotes the tuple $(x_{1},\hdots,x_{n})$, then we set $$\nabla_{\ov{x}\snull}h=\left[\frac{\partial h}{\partial x_{1}}\hdots\frac{\partial h}{\partial x_{n}}\right].$$ 

\begin{cons} Denote by $\ov{x}$ the tuple $(x_{1},\hdots,x_{n})$ and assume that $g\in\RR[\ov{x}]$ is a homogeneous polynomial of an even degree $d=\deg(g)$. We define six polynomials $f\in\RR[\ov{x}]$, $g_{\delta}\in\RR[\ov{x},\delta]$, $h_{0},h_{1}\in\RR[\ov{x},\delta]$ and $\wt{h}_{0},\wt{h}_{1}\in\RR[\ov{x},\delta,\gamma]$ as follows: $$f=\sum_{i=1}^{n}2^{i}x_{i}^{d},\quad g_{\delta}=(1-\delta)g+\delta(1+\sum_{j=1}^{n}x_{j}^{d}),$$ $$\textnormal{ }$$ $$h_{0}=\left\lVert \nabla_{\ov{x}\textnormal{ }}f\right\rVert^{2}=\sum_{i=1}^{n}(\frac{\partial f}{\partial x_{i}})^{2}=\sum_{i=1}^{n}2^{2i}d^{2}x_{i}^{2d-2},$$ $$h_{1}=\det(\left[\begin{array}{rr}\nabla_{\ov{x}\textnormal{ }}g_{\delta}\\\nabla_{\ov{x}\textnormal{ }}f\end{array}\right]\cdot\left[\begin{array}{rr}\nabla_{\ov{x}\textnormal{ }}g_{\delta}&\nabla_{\ov{x}\textnormal{ }}f\end{array}\right])+g^{2}_{\delta}=$$ $$=\det(\left[\begin{array}{rrrrrr}\frac{\partial g_{\delta}}{\partial x_{1}}&\hdots&\frac{\partial g_{\delta}}{\partial x_{n}}\\\\\frac{\partial f}{\partial x_{1}}&\hdots&\frac{\partial f}{\partial x_{n}}\end{array}\right]\cdot\left[\begin{array}{rrrrrr}\frac{\partial g_{\delta}}{\partial x_{1}}&\frac{\partial f}{\partial x_{1}}\\\vdots&\vdots\\\frac{\partial g_{\delta}}{\partial x_{n}}&\frac{\partial f}{\partial x_{n}}\end{array}\right])+g^{2}_{\delta}=\left|\begin{array}{rrrr}\sum_{i=1}^{n}(\frac{\partial g_{\delta}}{x_{i}})^{2}&\sum_{i=1}^{n}\frac{\partial g_{\delta}}{\partial x_{i}}\frac{\partial f}{\partial x_{i}}\\\\\sum_{i=1}^{n}\frac{\partial f}{\partial x_{i}}\frac{\partial g_{\delta}}{\partial x_{i}}&\sum_{i=1}^{n}(\frac{\partial f}{x_{i}})^{2}\end{array}\right|+g^{2}_{\delta},$$ $$\textnormal{  }$$ $$\wt{h_{0}}=(1-\gamma)h_{0}-(\sum_{i=1}^{n}x_{i}^{2d-2}\gamma),\quad \wt{h_{1}}=(1-\gamma)h_{1}-(\sum_{i=1}^{n}x_{i}^{4d-2}\gamma).$$ Observe that $\deg(\wt{h_{0}})=2d-1$ and $\deg(\wt{h_{1}})=4d-1$, because $\deg(h_{0})=2d-2$ and $\deg(h_{1})=4d-2$. Define $$\wh{h_{0i}}=\frac{\partial\wt{h_{0}}}{\partial x_{i}},\quad\wh{h_{1i}}=\frac{\partial\wt{h_{1}}}{\partial x_{i}},$$ for $i=1,\dots,n$ and view these polynomials as elements of the ring $\RR[\delta,\gamma][\ov{x}]$. Finally, set $$\mathcal{R}_{g}=\mathcal{R}(\wh{h_{01}},\dots,\wh{h_{0n}})\cup\mathcal{R}(\wh{h_{11}},\dots,\wh{h_{1n}})\subseteq\RR[u_{1},\dots,u_{n+1}]$$ where $\mathcal{R}(\wh{h_{01}},\dots,\wh{h_{0n}})$ and $\mathcal{R}(\wh{h_{11}},\dots,\wh{h_{1n}})$ are defined as in Construction 5.3. \epv
\end{cons}

\begin{remark} The above constructions are too tedious to give here any concrete examples. However, they are obviously computable and can be performed in any computer algebra system. \epv
\end{remark}

The following algorithm is a procedure for determining whether a given homogeneous polynomial $g\in\RR[x_{1},\dots,x_{n}]$ of an even degree satisfies the condition $g\geq 0$. We stick to the notation introduced earlier.

\begin{alg} Input: a homogeneous polynomial $g\in\RR[x_{1},\dots,x_{n}]$ of an even degree $d$. Output: \textit{yes} or \textit{no}, depending on the validity of the condition $g\geq 0$.

\begin{enumerate}[\rm(1)]
	\item Construct the set $\CR_{g}$, see Constructions 5.3 and 5.4 for details.
	\item Construct sets $J(n,d)$, $\CB(n,d)$ and calculate univariate polynomials $g_{j,\beta,r,-}^{un}$, $g_{j,\beta,r,+}^{un}$, $r_{n+1}$ as in Construction 5.1. Apply Algorithm 4.4 to determine whether the condition $$(\exists_{t}-g_{j,\beta,r,+}^{un}(t)>0\wedge r_{n+1}(t)>0)\vee(\exists_{t}-g_{j,\beta,r,-}^{un}(t)>0\wedge -r_{n+1}(t)>0)$$ does not hold, for any $j\in J(n,d)$, $\beta\in\CB(n,d)$ and $r\in\CR_{g}$. If this is the case, then the output is \textit{yes}. Otherwise, the output is \textit{no}.
\end{enumerate}
\end{alg}

{\bf Proof.} The procedure follows from Theorem 5.2. \epv

Algorithm 5.6 is very time-consuming from the point of view of computational complexity, partially because it relies on complicated constructions. However, this is the best known procedure for determining whether given homogeneous real polynomial of an even degree is non-negative. A detailed discussion of these issues is given in \cite{Re1}, see especially Section 1 and formulation of the main result presented as Proposition 4.2. 

The following procedure determines whether a hermiticity-preserving superoperator is positive. This is the main result of the paper.  

\begin{alg} Input: a hermiticity-preserving superoperator $\Phi\in\CL(\MM_{n}(\CCC))$. Output: \textit{yes} or \textit{no} depending on the positivity of $\Phi$.
\begin{enumerate}[\rm(1)]
	\item Calculate the positivity polynomial $p_{\Phi}$, see for example Remark 3.7.
	\item Apply Algorithm 5.6 to determine whether the condition $p_{\Phi}\geq 0$ holds. If this is the case, the output is \textit{yes}. Otherwise, the output is \textit{no}.
\end{enumerate}
\end{alg}

{\bf Proof.} The correctness of the algorithm follows directly from Algorithm 5.6 and Theorem 3.5. \epv

\section*{Acknowledgements} The authors are indebted to J. Renegar for his assistance in understanding the contents of \cite{Re1}.

\end{document}